\documentclass{article}
\usepackage{amssymb}
\usepackage{amsfonts}
\usepackage{amsmath}

\input{tcilatex}
\begin{document}

\title{Cosmology in $5D$\ and $4D$ Einstein-Gauss-Bonnet gravity}
\author{F. G\'{o}mez$^{1,a}$\thanks{%
fernagomez@udec.cl}, S. Lepe$^{2,b}$\thanks{%
samuel.lepe@pucv.cl}, V.C. Orozco$^{3,c}$\thanks{%
vorozco@udec.l} and P. Salgado$^{4,d}$\thanks{%
patsalgado@unap.cl} \\
$^{1}$Facultad de Ingenier\'{\i}a Agr\'{\i}cola, Universidad de Concepci\'{o}%
n, \\
Vicente Mendez 595, Chill\'{a}n, Chile.\\
$^{2}$Instituto de F\'{\i}sica, Pontificia Universidad Cat\'{o}lica de
Valpara\'{\i}so,\\
Avda. Brasil 2950, Valpara\'{\i}so, Chile.\\
$^{3}$Departamento de F\'{\i}sica, Universidad de Concepci\'{o}n, \\
Casilla 160-C, Concepci\'{o}n, Chile.\\
$^{4}$Instituto de Ciencias Exactas y Naturales, Facultad de Ciencias,\\
Universidad Arturo Prat, Avda. Arturo Prat 2120, Iquique, Chile.}
\maketitle

\begin{abstract}
We consider the five-dimensional Einstein-Gauss-Bonnet gravity, which can be
obtained by means of an apropriate choice of coefficients in the
five-dimensional Lanczos-Lovelock gravity theory\textbf{. \ }The
Einstein-Gauss-Bonnet field equations for the Friedmann-Lema\^{\i}%
tre-Robertson-Walker metric are found as well as some of their solutions.
The hyperbolicity of the corresponding equations of motion is discussed.

A four-dimensional gravity action is obtained from the Gauss-Bonnet gravity
using the Randall-Sundrum compactification procedure and then it is studied
the implications \ of the compactification procedure\ in the cosmological
solutions. The same procedure is used to obtain gravity in four dimensions
from the five-dimensional AdS-Chern-Simons gravity to then study some
cosmological solutions. Some aspects of the construction of the
four-dimensional action gravity, as well as a brief review of Lovelock
gravity in $5D$\ are considered in an Appendix.
\end{abstract}

\section{Introduction}

The $5$-dimensional action for Lanczos-Lovelock gravity theory \cite%
{Lovelock1,Zumino1986,TZ00,criso}, is a polynomial of degree $2$ in
curvature, which can be written in terms of the Riemann curvature $R^{ab}$
and the vielbein $e^{a}$ as 
\begin{equation}
S_{LL}^{(5D)}=\frac{1}{8\kappa _{5}}\int \varepsilon _{abcde}\left( \alpha
R^{ab}R^{cd}e^{e}+\frac{2}{3}R^{ab}e^{c}e^{d}e^{e}+\beta
\,e^{a}e^{b}e^{c}e^{d}e^{e}\right) ,  \label{uno}
\end{equation}%
where $\left( i\right) $ $\alpha ,\beta $ are arbitrary constants, $\left(
ii\right) $ $e^{a}=e_{\mu }^{a}\,\mathrm{d}x^{\mu },$ $\omega ^{ab}=\omega
_{\mu }^{\ ab}\,\mathrm{d}x^{\mu }$ are the f\"{u}nfbein fields and spin
connection, respectively, $\left( iii\right) $ $R^{ab}=\mathrm{d}\omega
^{ab}+\omega _{\ c}^{a}\omega ^{cb}$ is the $2$-form curvature and $\kappa
_{5}=12\pi ^{2}G_{5}$, where $G_{5}$ is the $5$-dimensional Newton constant.

Comparing the action (\ref{uno}), when $\alpha =0$, with the
Einstein-Hilbert-Cartan action with cosmological constant in $5D$%
\begin{equation}
S_{EHC}^{(5D)}=\frac{1}{12\kappa _{5}}\int \varepsilon _{abcde}\left(
R^{ab}e^{c}e^{d}e^{e}-\frac{\Lambda _{5D}}{10}\,e^{a}e^{b}e^{c}e^{d}e^{e}%
\right) ,  \label{dos}
\end{equation}%
we can see that the action (\ref{uno}) matches action (\ref{dos}) only if $%
\beta =-\Lambda _{5D}/15.$ With this choice of constant $\beta $, the action
(\ref{uno}) takes the form of Einstein-Gauss-Bonnet (EGB) action with
cosmological constant 
\begin{equation}
S_{EGB}^{(5D)}=\frac{1}{8\kappa _{5}}\int \varepsilon _{abcde}\left( \alpha
R^{ab}R^{cd}e^{e}+\frac{2}{3}R^{ab}e^{c}e^{d}e^{e}-\frac{\Lambda _{5D}}{15}%
\,e^{a}e^{b}e^{c}e^{d}e^{e}\right) .  \label{8}
\end{equation}%
In presence of matter, the action is given by 
\begin{equation}
S^{(5D)}=S_{EGB}^{(5D)}+S_{M}^{(5D)},  \label{9}
\end{equation}%
where $S_{M}^{(5D)}=S_{M}^{(5D)}(e^{a},\omega ^{ab})$ is the matter action
whose variation leads to 
\begin{equation}
\delta S_{M}^{(5D)}=\frac{\delta L_{M}^{(5D)}}{\delta e^{a}}\delta e^{a}\,+%
\frac{\delta L_{M}^{(5D)}}{\delta \omega ^{ab}}\delta \omega ^{ab}\,,
\label{os7}
\end{equation}%
where $\delta L_{M}^{(5D)}/\delta e^{a}$ and $\delta L_{M}^{(5D)}/\delta
\omega ^{ab}$ are related to the anholonomic forms (in an orthonormal frame)
of the energy-momentum tensor $T_{ab}$ and the spin tensor $S_{ab}^{c}$
respectively. This means that the variation of the action (\ref{9}) leads to
the following field equations 
\begin{align}
\varepsilon _{abcde}\left( \alpha R^{bc}R^{de}+2R^{bc}e^{d}e^{e}-\frac{%
\Lambda _{5D}}{3}e^{b}e^{c}e^{d}e^{e}\right) & =-8\kappa _{5}\frac{\delta
L_{M}^{(5D)}}{\delta e^{a}},  \label{os8} \\
2\varepsilon _{abcde}T^{c}\left( \alpha R^{de}+e^{d}e^{e}\right) & =-4\kappa
_{5}\frac{\delta L_{M}^{(5D)}}{\delta \omega ^{ab}},  \label{os9}
\end{align}%
where $T^{a}=\mathrm{D}e^{a}=\mathrm{d}e^{a}+\omega _{\ b}^{a}e^{b}$ is the $%
2$-form torsion. When the spin tensor is zero, one solution is the zero
torsion ($T^{a}=\mathrm{D}e^{a}=0$).

Summarizing, we have considered the $5$-dimensional Lanczos-Lovelock
gravity, which for an appropriate choice of coefficients leads to the EGB
gravity action\textbf{.\ }This work is organized as follows: In Section $2$
we find the EGB gravitational field equations for the Friedmann-Lema\^{\i}%
tre-Robertson-Walker (FLRW) metric. A discussion about the hyperbolicity of
the metric ends this section. Section $3$ is devoted to find a $4$%
-dimensional gravity action from the Gauss-Bonnet gravity using the
Randall-Sundrum compactification procedure and then we study the
implications in the cosmological solutions of the compactification
procedure. In Section $4$ we use the same procedure to obtain gravity in $4D$
from the $5$-dimensional AdS-Chern-Simons gravity and then we study some of
its cosmological implications. Finally Concluding Remarks are presented in
Section $5$. An appendix is included, where is considered a brief gravity
review of Lovelock gravity in\ $5D$, as well as some aspects of the
construction of the $4$-dimensional action gravity.

\section{Cosmology in Einstein-Gauss-Bonnet gravity without cosmological
constant}

Consider the action (\ref{9}) without cosmological constant, which means
that the lagrangian is given by%
\begin{equation}
L^{(5D)}=L_{EGB}^{(5)}|_{\Lambda _{5}=0}+L_{M}^{(5D)},  \label{4''}
\end{equation}%
with 
\begin{equation}
L_{EGB}^{(5)}\left( e,\omega \right) |_{\Lambda _{5}=0}=\frac{2}{3}%
\varepsilon _{abcde}R^{ab}e^{c}e^{d}e^{e}+\alpha \varepsilon
_{abcde}R^{ab}R^{cd}e^{e},  \label{3}
\end{equation}%
being $\alpha =\left\vert \alpha \right\vert sgn\left( \alpha \right) $ a
constant and $L_{M}^{(5D)}$ represents a matter lagrangian. Later we show
two cosmological scenarios associated with $sgn\left( \alpha \right) $.

The variation of the action $S^{(5D)}$ with respect to the vielbein $e^{a}$
and the spin connection $\omega ^{ab}$ leads to the equations%
\begin{eqnarray}
\varepsilon _{abcde}\left( 2R^{bc}e^{d}e^{e}+\left\vert \alpha \right\vert
sgn\left( \alpha \right) R^{bc}R^{de}\right) &=&-8\kappa _{5}\frac{\delta
L_{M}^{(5D)}}{\delta e^{a}},  \label{4'} \\
\varepsilon _{abcde}\left( T^{c}e^{d}e^{e}+\left\vert \alpha \right\vert
sgn\left( \alpha \right) R^{cd}T^{\text{ }e}\right) &=&-4\kappa _{5}\frac{%
\delta L_{M}^{(5D)}}{\delta \omega ^{ab}}.  \label{5'}
\end{eqnarray}%
If the matter under consideration has no spin, then%
\begin{equation}
\frac{\delta L_{M}}{\delta \omega ^{ab}}=0\text{ \ \ \ }and\text{\ \ \ \ }%
T^{a}=0.  \label{6'}
\end{equation}%
Since 
\begin{equation}
\frac{\delta L_{M}}{\delta e^{a}}=\frac{1}{4!}T_{a}^{\text{ \ }\mu
}\varepsilon _{\mu bcde}e^{b}e^{c}e^{d}e^{e},  \label{7}
\end{equation}%
where $T_{\mu \nu }$ is the energy-momentum tensor, we have that Eq. (\ref%
{4'}) takes the form%
\begin{equation}
\varepsilon _{abcde}\left( 2R^{bc}e^{d}e^{e}+\left\vert \alpha \right\vert
sgn\left( \alpha \right) R^{bc}R^{de}\right) =-\frac{\kappa _{5}}{3}T_{a}^{%
\text{ \ }\mu }\varepsilon _{\mu bcde}e^{b}e^{c}e^{d}e^{e}.  \label{8'}
\end{equation}

\subsection{Field equations and cosmology}

We consider a flat FLRW metric 
\begin{equation}
ds^{2}=-dt^{2}+a^{2}(t)\delta _{ij}dx^{i}dx^{j},  \label{9'}
\end{equation}%
wher $a(t)$ is the cosmic scale factor and $i,j=1,2,3,4$. After some
calculations, the $2$-form curvature turns out to be 
\begin{eqnarray}
R^{0p} &=&\frac{\ddot{a}}{a}e^{0}e^{p}=\left( \dot{H}+H^{2}\right)
e^{0}e^{p}=-qH^{2}e^{0}e^{p},\text{\ \ }  \label{10} \\
\text{\ }R^{pq} &=&H^{2}e^{p}e^{q},  \label{10'}
\end{eqnarray}%
where $p,q=1,2,3,4,$ $H=\dot{a}/a$ is the Hubble parameter, $\dot{H}=dH/dt=%
\ddot{a}/a-H^{2}$ and $q=-\left( 1+\dot{H}/H^{2}\right) $ is the
deceleration parameter. From here, it is direct to see that when $q<0$ we
have $\ddot{a}>0$ and if $q>0$ then $\ddot{a}<0.$

We further consider an energy-momentum tensor corresponding to a perfect
fluid 
\begin{equation}
T_{\mu }^{\text{ \ }\nu }=diag(-\rho ,p,p,p,p).  \label{11}
\end{equation}%
After replacing (\ref{10}) and (\ref{11}) in (\ref{8'}) we obtain the
Friedmann constraint and the conservation equation, respectively, 
\begin{equation}
6H^{2}+3\left\vert \alpha \right\vert sgn\left( \alpha \right) H^{4}=\kappa
_{5}\rho ,  \label{12}
\end{equation}%
\begin{equation}
\left( 1+z\right) \frac{d\rho }{dz}=4(\rho +p),  \label{13}
\end{equation}%
where we have introduced the redshift parameter defined as $1+z=a_{0}/a$ and 
$a_{0}=a\left( t_{0}\right) $. Choosing $\kappa _{5}=1$ unities and using
the barotropic equation of state $p=\omega \rho $, we write the equation (%
\ref{12}) and (\ref{13}) in the form

\begin{eqnarray}
6H^{2}\left( 1+\frac{1}{2}\left\vert \alpha \right\vert sgn\left( \alpha
\right) H^{2}\right) &=&\rho ,  \label{14} \\
\rho \left( z\right) &=&\rho \left( 0\right) \left( 1+z\right) ^{4\left(
1+\omega \right) },  \label{15}
\end{eqnarray}%
such that

\begin{eqnarray}
sgn\left( \alpha \right) &=&1\Longrightarrow H^{2}\left( z\right) =\frac{1}{%
\left\vert \alpha \right\vert }\left( \sqrt{1+\frac{\left\vert \alpha
\right\vert }{3}\rho \left( z\right) }-1\right) ,  \label{16} \\
sgn\left( \alpha \right) &=&-1\Longrightarrow H_{\pm }^{2}\left( z\right) =%
\frac{1}{\left\vert \alpha \right\vert }\left( 1\pm \sqrt{1-\frac{\left\vert
\alpha \right\vert }{3}\rho \left( z\right) }\right) .  \label{17}
\end{eqnarray}%
The last case shows an upper bound for $\rho $, that is, $\rho \left(
z_{s}\right) =3/\left\vert \alpha \right\vert $ and so

\begin{equation}
H_{\pm }\left( z_{s}\right) =\sqrt{\frac{1}{\left\vert \alpha \right\vert }},
\label{18}
\end{equation}%
being

\begin{equation}
z_{s}=-1+\left( \frac{3}{\left\vert \alpha \right\vert \rho \left( 0\right) }%
\right) ^{1/4\left( 1+\omega \right) }.  \label{19}
\end{equation}%
Replacing (\ref{19}) into (\ref{17}), we write

\begin{equation}
H_{\pm }^{2}\left( z\right) =\frac{1}{\left\vert \alpha \right\vert }\left(
1\pm \sqrt{1-\left( \frac{1+z}{1+z_{s}}\right) ^{4\left( 1+\omega \right) }}%
\right) ,  \label{20}
\end{equation}%
so that we have a solution for $H_{\pm }\left( z\right) $ if $z\leq z_{s}$.
According to (\ref{19}) and (\ref{20}) and considering $\omega =0$ (cold
dark matter), we have

\begin{equation}
z_{s}=-1+\left( \frac{3}{\left\vert \alpha \right\vert \rho \left( 0\right) }%
\right) ^{1/4}\gtreqqless 0.  \label{21}
\end{equation}%
If $-1<z_{s}<0$, we have $\rho \left( z\rightarrow z_{s}\right) \rightarrow
3/\left\vert \alpha \right\vert \Longrightarrow H_{\pm }\left( z\rightarrow
z_{s}\right) \rightarrow \sqrt{1/\left\vert \alpha \right\vert }$, i.e., a
future de Sitter evolution, unlike in $4D$-$\Lambda CDM$ where a de Sitter
evolution is reached when $z\rightarrow -1$.

If $z_{s}>0$, we have an unrealistic situation given that $\rho \left(
z\rightarrow z_{s}\right) \rightarrow 3/\left\vert \alpha \right\vert
\Longrightarrow H_{\pm }\left( z\rightarrow z_{s}\right) \rightarrow \sqrt{%
1/\left\vert \alpha \right\vert }$, that is, a past de Sitter evolution.

If $z_{s}=0$, we write

\begin{equation}
H_{\pm }^{2}\left( z\right) =\frac{\rho \left( 0\right) }{3}\left( 1\pm 
\sqrt{1-\left( 1+z\right) ^{4}}\right) ,  \label{22}
\end{equation}%
so that, $H_{\pm }^{2}\left( 0\right) =$ $\rho \left( 0\right) /3$ and $%
H_{+}^{2}\left( z\rightarrow -1\right) \rightarrow 2\rho \left( 0\right) /3$%
. Recalling that in the present discussion, see (\ref{3}), there is not
cosmological constant (thinking in a de Sitter evolution).

In the absence of $\rho $, from (\ref{17}) we obtain a self-accelerating
solution given by

\begin{equation}
H_{+}=\sqrt{\frac{2}{\left\vert \alpha \right\vert }}.  \label{23}
\end{equation}%
Substituting (\ref{15}) into (\ref{16}), it is straightforward to show that

\begin{equation}
\omega >-1\Longrightarrow \rho \left( z\rightarrow -1\right) \rightarrow 0%
\text{ \ \ }and\text{ \ \ }H\left( z\rightarrow -1\right) \rightarrow 0,
\label{24}
\end{equation}%
typical behavior of cosmic components not associated with dark energy.

We end this Section by highlighting what is shown in (\ref{17}), that is, an
upper bound for the present energy density $\rho \left( z_{s}\right)
=3/\left\vert \alpha \right\vert \longleftrightarrow H_{\pm }\left(
z_{s}\right) =\sqrt{1/\left\vert \alpha \right\vert }$. For comparison, in $%
4D$, the $\Lambda CDM$ model tell us that $\rho \left( z\rightarrow
-1\right) \rightarrow 0\Longrightarrow H\left( z\rightarrow -1\right)
\rightarrow \sqrt{\Lambda /3}$. We also highlight what is shown in (\ref{23}%
), a self-accelerating solution.

The deceleration parameter can be written as

\begin{equation}
q=-1+\frac{1}{2}\left( 1+z\right) \frac{dH^{2}}{H^{2}dz},  \label{24'}
\end{equation}%
that after using (\ref{14}-\ref{15}), it is straightforward to obtain

\begin{equation}
q\left( z\right) =-1+2\left( 1+\omega \right) f\left( z\right) ,  \label{32}
\end{equation}%
where we have defined

\begin{equation}
f\left( z\right) =\frac{1}{1+\alpha H^{2}\left( z\right) }\left( 1+\frac{1}{2%
}\alpha H^{2}\left( z\right) \right) .  \label{33}
\end{equation}%
and $f\left( z\right) <1$. The Friedmann constraint tells us that,
regardless of $sgn\left( \alpha \right) $, the following inequality must be
satisfied

\begin{equation}
1+\frac{1}{2}\alpha H^{2}\left( z\right) >0,  \label{34}
\end{equation}%
so that

\begin{equation}
q\left( z\right) <0\Longrightarrow 2\left( 1+\omega \right) <\frac{1}{%
f\left( z\right) }.  \label{35}
\end{equation}%
Considering $\omega \geq 0$ (fluids that by themselves generate
deceleration, $q=1+2\omega >0$), it is evident that $2\left( 1+\omega
\right) >1$, so that the inequality given in (\ref{35}) is satisfied.

In good accounts, it is perfectly possible to satisfy $q\left( z\right) <0$ (%
\ref{35}) so that we have consistency with the so-called Lorentzian metric
condition. In fact, according to Ref. \cite{sh1}, (see also \cite{sh3}-\cite%
{sh8}), "a necessary condition for hyperbolic EOM is that efective metric be
Lorentzian" and that the conditions for said metric to be Lorentzian is that

\begin{equation}
sgn\left( 1+4\lambda _{2}H^{2}\right) =sgn\left( 1-4\lambda
_{2}qH^{2}\right) ,  \label{81}
\end{equation}%
where now $\lambda _{2}=\alpha /4l_{p}^{3}$\ (see Appendix), as can be seen
from equations (3.11) and (3.12) of the aforementioned Ref. \cite{sh1}. In
this same reference it is established that if this equality is not
fulfilled, then the aforementioned metric will not be Lorentzian. This means
that if $\lambda _{2}>0$\ and $4\lambda _{2}qH^{2}>1$, we have a
non-Lorentzian metric. However, if $q<0$\ the metric is Lorentzian.

\section{Gravity in $4D$\ from Einstein-Gauss- Bonnet gravity}

The existence of new dimensions may have non trivial effects in our
understanding of the cosmology of the early Universe, among many other
issues. By convention, it has always been assumed that such extra dimensions
should be compactified to manifolds of small radii with sizes of the order
of the Planck length.

It was only in the last years of the $20th$ century when people started to
ask the question of how large could these extra dimensions be without
getting into conflict with observations. In this context, of particular
interest are the Randall and Sundrum models \cite{randall,randall1} for
warped backgrounds, with compact or even infinite extradimensions. Randall
and Sundrum proposed that the metric of the spacetime is given by

\begin{equation}
ds^{2}=e^{-2kr_{c}}\eta _{\mu \nu }dx^{\mu }dx^{\nu }+r_{c}^{2}d\phi ^{2},
\label{1}
\end{equation}%
i.e. a $4$-dimensional metric multiplied by a "warp\ factor" which is a
rapidly changing function of an additional dimension, $k$ is a scale of the
order of Planck scale, $x^{\mu }$ are coordinates for the familiar $4$%
-dimensions, while $0\leq \phi \leq \pi $ is the coordinate for an extra
dimension, which is a finite interval whose size is set by $r_{c}$, known as
"compactification radius".\ Randall and Sundrum showed that this metric is a
solution to Einstein's equations.

\subsection{4-dimensional gravity from the Einstein-Gauss-Bonnet gravity}

From equation (\ref{8}) we can see that the Lagrangian contains the
Gauss-Bonnet term, the Einstein-Hilbert term and a cosmological term.
Following the procedure given in the Appendix, we find that the $5$%
-dimensional action gravity compactified to $4$-dimensions is given by 
\begin{equation}
\tilde{S}[\tilde{e}]=\frac{1}{8\kappa _{5}}\int_{\Sigma _{4}}\tilde{%
\varepsilon}_{mnpq}\left( A\tilde{R}^{mn}\tilde{R}^{pq}+B\tilde{R}^{mn}%
\tilde{e}^{p}\tilde{e}^{q}+C\ \tilde{e}^{m}\tilde{e}^{n}\tilde{e}^{p}\tilde{e%
}^{q}\right) ,  \label{5t}
\end{equation}%
where%
\begin{eqnarray}
A &=&r_{c}\int_{0}^{2\pi }d\phi  \label{5t'} \\
B &=&2r_{c}\int_{0}^{2\pi }d\phi e^{2f(\phi )}\left( 1-\frac{\alpha }{%
r_{c}^{2}}\left( 3f^{\prime 2}+2f^{\prime \prime }\right) \right) ,
\label{5t''}
\end{eqnarray}%
and%
\begin{equation}
C=r_{c}\int_{0}^{2\pi }d\phi e^{4f(\phi )}\left( \frac{\alpha }{r_{c}^{4}}%
f^{\prime 2}\left( 5f^{\prime 2}+4f^{\prime \prime }\right) -\frac{2}{%
3r_{c}^{2}}\left( 5f^{\prime 2}+2f^{\prime \prime }\right) -\frac{\Lambda
_{5D}}{3}\right) .  \label{5t'''}
\end{equation}%
Since $f(\phi )$ is arbitrary and continuously differentiable function, and
since we are working with a cylindrical variety, we find that (\ref{5t'}), (%
\ref{5t''}) lead to

\begin{eqnarray}
A &=&2\pi r_{c}  \label{6t} \\
B &=&2\pi r_{c}\left( 1+\frac{\alpha }{r_{c}^{2}}\right) ,  \label{6t1}
\end{eqnarray}%
and%
\begin{equation}
B=-\frac{\pi }{4r_{c}}\left( \frac{\alpha }{r_{c}^{2}}-2+\Lambda
_{5D}r_{c}^{2}\right) ,  \label{6t'}
\end{equation}%
were we have choose $f(\phi )=\ln \left( \sin \phi \right) $.

Note that in the action (\ref{5t}) there is a quadratic term in the
curvature given by $A\tilde{\varepsilon}_{mnpq}\tilde{R}^{mn}\tilde{R}^{pq}$%
, which represents the $4$-dimensional Gauss-Bonnet term. This term is a
topological one, so that it does not contribute to the dynamics and it can
be eliminated. \ This means that compactification avoids the problems cited
in Ref. \cite{sh2} (see also \cite{sh9}-\cite{sh12}). Equation $\left(
1\right) $\ of this reference agrees with the Lagrangian (\ref{5t}), except
for the cosmological term, when $\lambda _{1}=\alpha $, $\lambda
_{2}/\lambda _{1}=-4$\ and $\lambda _{3}/\lambda _{1}=1.$

Taking into account that the action (\ref{5t}) should lead to the
four-dimensional Einstein-Hilbert-Cartan action, namely%
\begin{equation}
\tilde{S}_{EHC}^{(4D)}=\frac{1}{4\kappa _{4}}\int_{\Sigma _{4}}\tilde{%
\varepsilon}_{mnpq}\left( \tilde{R}^{mn}\tilde{e}^{p}\tilde{e}^{q}-\frac{%
\Lambda _{4D}}{6}\ \tilde{e}^{m}\tilde{e}^{n}\tilde{e}^{p}\tilde{e}%
^{q}\right) ,  \label{6t''}
\end{equation}%
where $\kappa _{4}=8\pi G,$ it is direct to see that this occurs when 
\begin{equation}
B=3\pi \frac{G_{5}}{G}\text{ \ \ },\text{ \ \ }C=-\frac{\pi }{2}\Lambda _{4D}%
\frac{G_{5}}{G}.  \label{7t}
\end{equation}%
On the other hand we know that if $G_{D}$ is Newton's constant in $D$%
-dimensions and if $G$ is the usual Newton's constant, then 
\begin{equation*}
G_{D}=\left( l_{C}\right) ^{D-4}G,
\end{equation*}%
where $l_{C}$ is the length of the extra compact dimension \cite{zwiebach}.
In our particular case, $D=5$ and then $l_{C}=2\pi r_{c}$. This means that $%
G_{5}=2\pi r_{c}G$. So that (\ref{7t}) takes the form%
\begin{equation}
B=6\pi ^{2}r_{c}\text{ \ \ },\text{ \ \ }C=-\pi ^{2}\Lambda _{4D}r_{c}.
\label{7t'}
\end{equation}%
Now, from (\ref{6t}), (\ref{6t'}) and (\ref{7t'}), it is direct to see that

\begin{equation}
\frac{\alpha }{r_{c}^{2}}=3\pi -1,  \label{24t}
\end{equation}%
and then%
\begin{equation}
\Lambda _{4D}=\Lambda _{4D}\left( r_{c},\Lambda _{5D}\right) =\frac{1}{4\pi }%
\left( \Lambda _{5D}+\frac{3(\pi -1)}{r_{c}^{2}}\right) .  \label{25t}
\end{equation}%
Introducing (\ref{7t'}) into (\ref{5t}) we obtain the action (\ref{6t''})
where now $\Lambda _{4D}$ is given by (\ref{25t}).

In tensor language the two terms in (\ref{5t}) can be written as%
\begin{eqnarray}
\tilde{\varepsilon}_{mnpq}\tilde{R}^{mn}\tilde{e}^{p}\tilde{e}^{q} &=&-2\int
d^{4}\tilde{x}\sqrt{-\tilde{g}}\tilde{R},  \notag \\
\tilde{\varepsilon}_{mnpq}\tilde{e}^{m}\tilde{e}^{n}\tilde{e}^{p}\tilde{e}%
^{q} &=&-24\int d^{4}\tilde{x}\sqrt{-\tilde{g}},  \label{29t}
\end{eqnarray}%
where $\tilde{g}$ is the determinant of the $4$-dimensional metric tensor $%
\tilde{g}_{\mu \nu }$ and $\tilde{R}$ is the Ricci scalar. Thus, the action (%
\ref{5t}) is now written as

\begin{equation}
\tilde{S}[\tilde{g}]=\int d^{4}\tilde{x}\sqrt{-\tilde{g}}\left( \tilde{R}%
+2\Lambda _{4D}\right) ,  \label{31t}
\end{equation}%
whose field equations are 
\begin{equation}
G_{\mu \nu }=\Lambda _{4D}g_{\mu \nu },
\end{equation}%
and $\Lambda _{4D}=\Lambda _{4D}\left( r_{c},\Lambda _{5D}\right) $.

According to (40), we can say little or nothing about the presence of $r_{c}$%
. The only thing we can "speculate" is to say that $\Lambda _{4D}$
originates from the compactification radius and the $5$-dimensional
cosmological constant\textbf{, }and nothing else.

\section{\textbf{Cosmology in AdS Chern-Simons gravity compactified to
4-dimensions}}

Consider again the EGB action (\ref{8}). Choosing $\alpha =l^{2}$ and $%
\Lambda =-3/l^{2}$ in (\ref{8}), we see that the EGB action takes the form 
\begin{equation}
S_{EGB}=\frac{l^{2}}{8\kappa _{5}}\int \varepsilon _{abcde}\left(
R^{ab}R^{cd}e^{e}+\frac{2}{3l^{2}}R^{ab}e^{c}e^{d}e^{e}+\frac{1}{5l^{4}}%
\,e^{a}e^{b}e^{c}e^{d}e^{e}\right) ,  \label{ap3}
\end{equation}%
where it is straightforward to see that this particular choice for $\alpha $
and $\Lambda $ in the EGB action leads to the $5$-dimensional Chern-Simons
gravity action for the AdS algebra, with $l$ interpreted as the radius of
the universe.

\subsection{\textbf{4-dimensional gravity from the AdS Chern-Simons gravity }%
}

From equation (\ref{ap3}) we can see that the Lagrangian contains the
Gauss-Bonnet term $L_{GB}$, the Einstein-Hilbert term $L_{EH}$ and a
cosmological term $L_{\Lambda }$. Replacing (\ref{4t'}), (\ref{4t''}) and (%
\ref{4t'''}) in (\ref{ap3}) we find 
\begin{equation}
\tilde{S}[\tilde{e}]=\frac{1}{8\kappa _{5}}\int_{\Sigma _{4}}\tilde{%
\varepsilon}_{mnpq}\left( \tilde{A}\tilde{R}^{mn}\tilde{e}^{p}\tilde{e}^{q}+%
\tilde{B}\ \tilde{e}^{m}\tilde{e}^{n}\tilde{e}^{p}\tilde{e}^{q}\right) ,
\label{999}
\end{equation}%
where%
\begin{equation}
\tilde{A}=\frac{2\pi l^{2}}{r_{c}}\left( 1+\frac{r_{c}^{2}}{l^{2}}\right) ,
\label{20t}
\end{equation}%
and%
\begin{equation}
\tilde{B}=-\frac{\pi }{4r_{c}}\left( \frac{l^{2}}{r_{c}^{2}}-2-3\frac{%
r_{c}^{2}}{l^{2}}\right) .  \label{21t}
\end{equation}%
It is direct to see that the action (\ref{999}) lead to the
Einstein-Hilbert-Cartan action when 
\begin{equation}
\tilde{A}=6\pi ^{2}r_{c}\text{ \ \ }and\text{ \ \ }\tilde{B}=-\pi
^{2}\Lambda _{4D}r_{c}.  \label{21t'}
\end{equation}%
From (\ref{20t}), (\ref{21t}) and (\ref{21t'}) we have

\begin{equation}
\frac{r_{c}^{2}}{l^{2}}=\frac{1}{3\pi -1},  \label{24t'}
\end{equation}%
and then \ 
\begin{equation}
\text{ }\Lambda _{4D}=\Lambda _{4D}\left( r_{c}\right) =\left( \frac{3\pi -4%
}{3\pi -1}\right) \frac{3}{4r_{c}^{2}}.  \label{25tc}
\end{equation}%
The introduction of (\ref{21t'}) into the action (\ref{999}) leads to the
action (\ref{6t''}) where now $\Lambda _{4D}$ is given by (\ref{25tc}).

Introducing (\ref{29t}) in (\ref{999}) we obtain

\begin{equation}
\tilde{S}[\tilde{g}]=\int d^{4}\tilde{x}\sqrt{-\tilde{g}}\left( \tilde{R}%
-2\Lambda _{4D}\right) ,  \label{31t'}
\end{equation}%
whose field equations are 
\begin{equation}
G_{\mu \nu }=-\Lambda _{4D}g_{\mu \nu }.
\end{equation}

In order to have a feeling on $r_{c}$, from (\ref{25tc}) we obtain $%
r_{c}\approx \Lambda _{4D}^{-1/2}$. Using $\Lambda _{4D}\sim
10^{-52}[m^{-2}]\sim 3\ast 10^{-122}[l_{Planck}^{-2}]$ we have $r_{c}\sim
10^{61}[l_{Planck}]\approx 10^{26}[m]$, we recalling that $a_{0}\approx
10^{26}[m]$ (current causal size of the universe).

According to (\ref{21t'}), we obtain $r_{c}/l\approx 0.34$, i.e., $l\approx $
$3r_{c}$. So, interpreting $l$\ \textit{as} \textit{the size of the universe}
appears to be reasonable.

\section{\textbf{Concluding remarks}}

We have considered the $5$-dimensional Lanczos-Lovelock gravity, which for
an appropriate choice of coefficients gives the EGB gravity action\textbf{. }%
It is found the EGB gravitational field equations for the FLRW metric
together with some cosmological solutions. \ And if the deceleration
parameter is negative, the so-called Lorentzian metric condition is
satisfied (see \cite{sh1}).

The main purpose of this article was to make the $5$-dimensional EGB gravity
theory, as well as the $5$-dimensional AdS-Chern-Simons, consistent with the
idea of a $4$-dimensional spacetime, through the replacement of a
Randall-Sundrum type metric in the Lagrangian (\ref{8}), and then\ to get an
interpretation of the $4$-dimensional effective cosmological constant.

We have evaluated a $5$-dimensional Randall-Sundrum type metric in the
Lagrangians (\ref{8}) and (\ref{ap3}), and then we derive an action for a $4$%
-dimensional spacetime embedded in the $5$-dimensional spacetime. We have
obtained the actions in tensorial language and then we find the
corresponding Friedmann equations for homogeneous and isotropic cosmology.

The quadratic term in the curvature of the action (\ref{5t}) given by $A%
\tilde{\varepsilon}_{mnpq}\tilde{R}^{mn}\tilde{R}^{pq}$\ represents the $4$%
-dimensional Gauss-Bonnet term. This term is a topological one, so that it
does not contribute to the dynamics. \ This means that compactification
avoids the problems cited in Ref. \cite{sh2}. Equation $\left( 1\right) $\
of this last reference agrees with the Lagrangian (\ref{5t}), except for the
cosmological term, when $\lambda _{1}=\alpha $, $\lambda _{2}/\lambda
_{1}=-4 $\ and $\lambda _{3}/\lambda _{1}=1.$\ 

Finally, it is important to note that the equations of motion corresponding
both the action (\ref{3}) and the action (\ref{5t}) are second order, so
they do not experience instabilities (see details in Ref. \cite{sh2}).

\textbf{Acknowledgements: }This work was supported in part by\textit{\ }%
FONDECYT Grants\textit{\ }No.\textit{\ }1180681 and No.\textit{\ }1211219
from the Government of Chile. One of the authors (VCO) was supported by
Universidad de Concepci\'{o}n\textit{,} Chile.

\section{\textbf{Appendix: }A briefly review the of derivation of the action%
\textbf{\ }(\protect\ref{5t}) and of lovelock gravity in\textbf{\ }$5D$}

\subsection{Gravity in $4D$ from EGB gravity}

In order to find (\ref{5t}), we will first consider the following $5$%
-dimensional Randall-Sundrum type metric \cite{rd}%
\begin{eqnarray}
ds^{2} &=&e^{2f(\phi )}\tilde{g}_{\mu \nu }(\tilde{x})d\tilde{x}^{\mu }d%
\tilde{x}^{\nu }+r_{c}^{2}d\phi ^{2},  \notag \\
&=&\eta _{ab}e^{a}e^{b},  \notag \\
&=&e^{2f(\phi )}\tilde{\eta}_{mn}\tilde{e}^{m}\tilde{e}^{n}+r_{c}^{2}d\phi
^{2},  \label{5t2}
\end{eqnarray}%
where $e^{2f(\phi )}$ is the so-called "warp factor", and $r_{c}$ is the
so-called "compactification radius" of the extra dimension, which is
associated with the coordinate $0\leqslant \phi <2\pi $. The symbol $\sim $
denotes $4$-dimensional quantities. We will use the usual notation \cite{rd}%
, \cite{gomez} 
\begin{eqnarray}
x^{\alpha } &=&\left( \tilde{x}^{\mu },\phi \right) \text{ \ \ };\text{ \ \ }%
\alpha ,\beta =0,...,4\text{ \ \ };\text{ \ \ }a,b=0,...,4;  \notag \\
\mu ,\nu &=&0,...,3\text{ \ \ };\text{ \ \ }m,n=0,...,3\text{ };  \notag \\
\eta _{ab} &=&diag(-1,1,1,1,1)\text{ \ \ };\text{ \ \ }\tilde{\eta}%
_{mn}=diag(-1,1,1,1),  \label{6t2}
\end{eqnarray}%
which allows us to write the vielbein 
\begin{equation}
e^{m}(\phi ,\tilde{x})=e^{f(\phi )}\tilde{e}^{m}(\tilde{x})=e^{f(\phi )}%
\tilde{e}_{\text{ }\mu }^{m}(\tilde{x})d\tilde{x}^{\mu }\text{ \ \ };\text{
\ \ }e^{4}(\phi )=r_{c}d\phi ,\text{\ }  \label{s22}
\end{equation}%
where\textbf{\ }$\tilde{e}^{m}$\textbf{\ }is the vierbein.

From the vanishing torsion condition%
\begin{equation}
T^{a}=de^{a}+\omega _{\text{ }b}^{a}e^{b}=0,  \label{2t}
\end{equation}%
we obtain the connections 
\begin{equation}
\omega _{\text{ }b\alpha }^{a}=-e_{\text{ }b}^{\beta }\left( \partial
_{\alpha }e_{\text{ }\beta }^{a}-\Gamma _{\text{ }\alpha \beta }^{\gamma }e_{%
\text{ }\gamma }^{a}\right) ,  \label{3t}
\end{equation}%
where $\Gamma _{\text{ }\alpha \beta }^{\gamma }$ is the Christoffel symbol.

From Eqs. (\ref{s22}) and (\ref{2t}) we find%
\begin{equation}
\omega _{\text{ }4}^{m}=\frac{e^{f}f^{\prime }}{r_{c}}\tilde{e}^{m},
\label{102t}
\end{equation}%
and the $4$-dimensional vanishing torsion condition 
\begin{equation}
\tilde{T}^{m}=\tilde{d}\tilde{e}^{m}+\tilde{\omega}_{\text{ }n}^{m}\tilde{e}%
^{n}=0,  \label{1030t}
\end{equation}%
where\textbf{\ \ }$f^{\prime }=\partial f/\partial \phi $\textbf{, }$\tilde{%
\omega}_{\text{ }n}^{m}=\omega _{\text{ }n}^{m}$\textbf{\ }and\textbf{\ }$%
\tilde{d}=d\tilde{x}^{\mu }\partial /\partial \tilde{x}^{\mu }$.

From (\ref{102t}), (\ref{1030t}) and the Cartan's second structural
equation, $R^{ab}=d\omega ^{ab}+\omega _{\text{ }c}^{a}\omega ^{cb}$, we
obtain the components of the $2$-form curvature \cite{rd}, \cite{gomez}%
\begin{equation}
R^{m4}=\frac{e^{f}}{r_{c}}\left( f^{\prime 2}+f^{\prime \prime }\right)
d\phi \tilde{e}^{m},\text{ \ }R^{mn}=\tilde{R}^{mn}-\left( \frac{%
e^{f}f^{\prime }}{r_{c}}\right) ^{2}\tilde{e}^{m}\tilde{e}^{n},\text{\ }
\label{105t}
\end{equation}%
where the $4$-dimensional $2$-form curvature is given by%
\begin{equation}
\tilde{R}^{mn}=\tilde{d}\tilde{\omega}^{mn}+\tilde{\omega}_{\text{ }p}^{m}%
\tilde{\omega}^{pn}.
\end{equation}%
These results allow us to obtain an action for a $4$-dimensional gravity
from the $5$-dimensional EGB action with cosmological constant, whose action
is given by (\ref{8}).

From (\ref{8}) we can see that the Lagrangian contains the Gauss-Bonnet term 
$L_{GB}$, the Einstein-Hilbert term $L_{EH}$ and a cosmological term $%
L_{\Lambda }$. In fact, replacing (\ref{s22}) and (\ref{105t}) in $L_{GB},$ $%
L_{EH},$ $L_{\Lambda }$ and using $\tilde{\varepsilon}_{mnpq}=\varepsilon
_{mnpq4}$, we obtain

\begin{eqnarray}
L_{GB} &=&\varepsilon _{abcde}R^{ab}R^{cd}e^{e},  \notag \\
&=&r_{c}d\phi \left\{ \tilde{\varepsilon}_{mnpq}\tilde{R}^{mn}\tilde{R}%
^{pq}-\left( \frac{2e^{2f}}{r_{c}^{2}}\right) \left( 3f^{\prime
2}+2f^{\prime \prime }\right) \tilde{\varepsilon}_{mnpq}\tilde{R}^{mn}\tilde{%
e}^{p}\tilde{e}^{q}\right.  \notag \\
&&\text{ \ \ \ \ \ \ \ \ \ \ \ \ \ \ }\left. +\left( \frac{e^{4f}}{r_{c}^{4}}%
f^{\prime 2}\right) \left( 5f^{\prime 2}+4f^{\prime \prime }\right) \tilde{%
\varepsilon}_{mnpq}\tilde{e}^{m}\tilde{e}^{n}\tilde{e}^{p}\tilde{e}%
^{q}\right\} ,  \label{4t'}
\end{eqnarray}

\begin{eqnarray}
L_{EH} &=&\varepsilon _{abcde}R^{ab}e^{c}e^{d}e^{e},  \notag \\
&=&r_{c}d\phi \left\{ \left( 3e^{2f}\right) \tilde{\varepsilon}_{mnpq}\tilde{%
R}^{mn}\tilde{e}^{p}\tilde{e}^{q}\right.  \notag \\
&&-\left. \left( \frac{e^{4f}}{r_{c}^{2}}\right) \left( 5f^{\prime
2}+2f^{\prime \prime }\right) \tilde{\varepsilon}_{mnpq}\tilde{e}^{m}\tilde{e%
}^{n}\tilde{e}^{p}\tilde{e}^{q}\right\} ,  \label{4t''}
\end{eqnarray}%
and

\begin{eqnarray}
L_{\Lambda } &=&\varepsilon _{abcde}e^{a}e^{b}e^{c}e^{d}e^{e},  \notag \\
&=&5r_{c}d\phi e^{4f}\tilde{\varepsilon}_{mnpq}\tilde{e}^{m}\tilde{e}^{n}%
\tilde{e}^{p}\tilde{e}^{q}.  \label{4t'''}
\end{eqnarray}

\subsection{Lovelock gravity in\textbf{\ }$5D$}

En las ecuaciones $\left( 2.1\right) $ y $\left( 2.2\right) $ de la Ref. 
\cite{sh1} the coefficients $\lambda _{k}$ in the Lagrangian $\left(
2.2\right) $ have dimensions of [length]$^{\left( 2p-D\right) }$ and $\delta
_{j_{1}\cdot \cdot \cdot j_{2p}}^{i_{1}\cdot \cdot \cdot i_{2p}}$ are the
so-called generalized Kronecker delta. Usually such Lagrangian density is
normalized in units of Planck length $\lambda _{1}=\left( 16\pi G\right)
^{-1}=l_{P}^{2-D}.$ In $5$-dimensions, the Lagrangian is given by the first
three terms of the sum

\begin{equation}
\mathcal{L}^{(5D)}=\sqrt{-g}\left[ \lambda _{0}+\lambda _{1}R+\lambda
_{2}\left( R^{2}-4R_{ij}R^{ij}+R_{ijkl}R^{ijkl}\right) \right] ,  \label{a4'}
\end{equation}%
where $\lambda _{1}=\left( 16\pi G\right) ^{-1}=l_{P}^{-3}$.

In the language of differential forms, the five-dimensional Lovelock
Lagrangian can be written as \cite{criso}

\begin{equation}
\mathcal{L}^{(5)}=\varepsilon _{abcde}\left( \alpha
_{0}\,e^{a}e^{b}e^{c}e^{d}e^{e}+\alpha _{1}R^{ab}e^{c}e^{d}e^{e}+\alpha
_{2}R^{ab}R^{cd}e^{e}\right) ,  \label{a7}
\end{equation}%
where $\alpha _{1}$, $\alpha _{2}$ and $\alpha _{3}$ are arbitrary constants.

Taking into account that $\varepsilon _{abcde}e^{a}e^{b}e^{c}e^{d}e^{e}=-120%
\sqrt{-g}d^{5}x$, $\varepsilon _{abcde}R^{ab}e^{c}e^{d}e^{e}=-6\sqrt{-g}%
Rd^{5}x$, $\varepsilon _{abcde}R^{ab}R^{cd}e^{e}=-\sqrt{-g}\left(
R^{2}-4R_{ij}R^{ij}+R_{ijkl}R^{ijkl}\right) d^{5}x$, we have that (\ref{a7})
can be written in the form

\begin{equation}
\mathcal{L}^{(5)}=-\sqrt{-g}\left( 120\alpha _{0}+6\alpha _{1}R+\alpha _{2}%
\left[ R^{2}-4R_{ij}R^{ij}+R_{ijkl}R^{ijkl}\right] \right) d^{5}x.
\label{a9}
\end{equation}%
The comparison of (\ref{a4'}) with (\ref{a9}) we see that $\lambda
_{0}=120\alpha _{0}$, $\lambda _{1}=6\alpha _{1}$, $\lambda _{2}=\alpha _{2}$%
.

On the another hand, from (\ref{8}) it is direct to see

\begin{equation}
\mathcal{L}_{EGB}^{(5D)}=\varepsilon _{abcde}\left( \alpha R^{ab}R^{cd}e^{e}+%
\frac{2}{3}R^{ab}e^{c}e^{d}e^{e}-\frac{\Lambda _{5D}}{15}%
\,e^{a}e^{b}e^{c}e^{d}e^{e}\right) ,  \label{a10}
\end{equation}%
where $\alpha =2\alpha _{2}/3\alpha _{1}$, $\beta =2\alpha _{0}/3\alpha _{1}$%
, which indicates that the coefficients $\alpha $ and $\lambda _{2}$ are
proportional. Indeed

\begin{equation}
\alpha =\frac{2\alpha _{2}}{3\alpha _{1}}=\frac{4\lambda _{2}}{\lambda _{1}}%
=64\pi G\lambda _{2}=4l_{P}^{3}\lambda _{2}  \label{a11}
\end{equation}

\end{document}